\documentclass[aps,prl,reprint,preprintnumbers,superscriptaddress,nofootinbib,longbibliography,floatfix]{revtex4-1}

\usepackage[utf8]{inputenc}
\usepackage[T1]{fontenc}
\usepackage{lmodern}
\usepackage[toc,page]{appendix}
\usepackage{microtype}
\usepackage{mathtools}

\usepackage{placeins}
\usepackage[caption=false]{subfig}
\usepackage{amsfonts}
\usepackage{hyperref}
\hypersetup{
  colorlinks=true,
  citecolor=blue,
  linkcolor=blue,
  urlcolor=blue
}

\usepackage{verbatim}
\usepackage{amsmath}
\usepackage{physics}
\usepackage{amssymb}
\usepackage{multirow}
\usepackage{slashed}

\usepackage{color}
\definecolor{mit-red}{rgb}{0.64,.12,0.2}
\definecolor{darkred}{rgb}{1.0,0.1,0.1}
\definecolor{darkgreen}{rgb}{0.1,0.7,0.1}
\definecolor{darkblue}{rgb}{0.1,0.1,1.0}
\definecolor{pink}{rgb}{1.0, 0.0, 0.67}

\usepackage{amsmath}

\DeclareRobustCommand{\Fig}[1]{Fig.~\ref{fig:#1}}
\DeclareRobustCommand{\Figs}[2]{Figs.~\ref{fig:#1} and \ref{fig:#2}}
\DeclareRobustCommand{\Eq}[1]{Eq.~(\ref{eq:#1})}

\DeclareRobustCommand{\Reference}[1]{Ref.~\cite{#1}}

\def\GeV{\text{GeV}}
\def\Cathode{\textsc{CATHODE}}

\begin{document}

\title{Isolating Unisolated Upsilons with Anomaly Detection in CMS Open Data}

\preprint{MIT-CTP 5843}

\author{Rikab Gambhir}
\email{rikab@mit.edu}
\affiliation{Center for Theoretical Physics, Massachusetts Institute of Technology, Cambridge, MA 02139, USA}
\affiliation{The NSF AI Institute for Artificial Intelligence and Fundamental Interactions}

\author{Radha Mastandrea}
\email{rmastand@berkeley.edu}
\affiliation{Department of Physics, University of California, Berkeley, CA 94720, USA}
\affiliation{Physics Division, Lawrence Berkeley National Laboratory, Berkeley, CA 94720, USA}

\author{Benjamin Nachman}
\email{bpnachman@lbl.gov}
\affiliation{Physics Division, Lawrence Berkeley National Laboratory, Berkeley, CA 94720, USA}
\affiliation{Berkeley Institute for Data Science, University of California, Berkeley, CA 94720, USA}

\author{Jesse Thaler}
\email{jthaler@mit.edu}
\affiliation{Center for Theoretical Physics, Massachusetts Institute of Technology, Cambridge, MA 02139, USA}
\affiliation{The NSF AI Institute for Artificial Intelligence and Fundamental Interactions}

\begin{abstract}

We present the first study of anti-isolated Upsilon decays to two muons ($\Upsilon \to \mu^+ \mu^-$) in proton-proton collisions at the Large Hadron Collider.
Using a machine learning (ML)-based anomaly detection strategy, we ``rediscover'' the $\Upsilon$ in 13 TeV CMS Open Data from 2016, despite overwhelming anti-isolated backgrounds.
We elevate the signal significance to $6.4 \sigma$ using these methods, starting from $1.6 \sigma$ using the dimuon mass spectrum alone.
Moreover, we demonstrate improved sensitivity from using an ML-based estimate of the multi-feature likelihood compared to traditional ``cut-and-count'' methods.
Our work demonstrates that it is possible and practical to find real signals in experimental collider data using ML-based anomaly detection, and we distill a readily-accessible benchmark dataset from the CMS Open Data to facilitate future anomaly detection developments.
\end{abstract}

\maketitle

Quarkonia fragmentation within jets is an important test of quantum chromodynamics (QCD), since it probes the transition between perturbative and nonperturbative scales~\cite{Ernstrom:1996am, Andronic:2015wma, Bain:2016clc, Celiberto:2022dyf}.
At the Large Hadron Collider (LHC), measurements of charmonium (e.g.~$J/\psi$) production within jets by LHCb~\cite{LHCb:2017llq} and CMS~\cite{CMS:2021puf} have led to improved fragmentation models~\cite{Bain:2017wvk} compared to baseline parton shower predictions with leading-order fragmentation functions~\cite{Bodwin:1994jh}.
To date, the only study of bottomonium (e.g.~$\Upsilon$) inside jets is an LHCb thesis that emphasized relatively isolated $\Upsilon$'s~\cite{Cooke:2023ukz}.
To our knowledge, no published study has looked for \emph{anti-isolated} $\Upsilon$ resonances, whose decay products are \emph{not} well-separated from other collision activity, thereby offering a complementary probe of QCD fragmentation.
Unlike anti-isolated $J/\psi$'s, whose signature in the dimuon channel is readily visible over the background, anti-isolated $\Upsilon$'s are rarer and are more difficult to identify through simple event selections.
This motivates the use of high-dimensional auxiliary features to reveal anti-isolated $\Upsilon$'s in collider data.

In this paper, we present the first study of anti-isolated $\Upsilon \to \mu^+ \mu^-$ decays at the LHC.
Our analysis is based on $13$ TeV proton-proton collision data taken in 2016, as accessed via the CMS Open Data \cite{opendata}.
By applying machine learning (ML)-based \textit{anomaly detection}, we identify a statistically significant sample of anti-isolated $\Upsilon$'s, elevating a marginal $1.6\sigma$ excess to well over the 5$\sigma$ discovery threshold.
The anomaly detection method we use, called CATHODE~\cite{Hallin:2021wme}, is capable of detecting resonant features without requiring a simulation of the signal or background.
We further show how to boost the signal significance by reweighting the data according to the learned multi-feature likelihood, which yields better sensitivity than a simple ``cut-and-count'' on the anomaly score.

The discovery potential of ML-based anomaly detection has been amply demonstrated on synthetic datasets for the LHC Olympics~\cite{Kasieczka:2021xcg} and the Dark Machine Anomaly Score Challenge \cite{Aarrestad:2021oeb}; see \Reference{livingReview} for a review of related techniques.
At the LHC, CMS and ATLAS have started to apply ML-based anomaly detection to search for new physics in jet data, though no significant excesses have emerged~\cite{ATLAS:2020iwa, ATLAS:2023azi, CMS:2024nsz, ATLAS:2023ixc, ATLAS:2025obc}.
Past work on CMS Open Data used ML-based anomaly detection to rediscover the top quark in already-known channels~\cite{knapp2020adversarially}. 
Our study of anti-isolated $\Upsilon$'s builds on this literature by showing that ML-based anomaly detection is capable of finding real signals in real experimental data in regions of phase space not previously studied.
To facilitate further ML method development and complement existing synthetic benchmarks, we publish a curated slice of the CMS Open Data used in this study along with the code to reproduce our analysis.


\textbf{\textit{Anti-Isolated Upsilon Selection.}}
The $\Upsilon$'s are a series of spin-1 bottom-antibottom ($b\Bar{b}$) resonances with masses $m_{\Upsilon} \geq 2m_b \approx 10 $ GeV~\cite{ParticleDataGroup:2024cfk}.
There are three main resonances with substantial dimuon branching ratios of $\approx 2\%$:  $\Upsilon(1S)$, $\Upsilon(2S)$, and $\Upsilon(3S)$.
(The $\Upsilon(4S)$, well-known for its role in $B$ factory physics, has a thousand-fold smaller dimuon branching ratio.)
Our analysis is focused on resonant anti-isolated $\Upsilon \to \mu^+\mu^-$ decays as seen by CMS in 2016 at $\sqrt{s} = 13$ TeV. 
This data is made public through the CERN Open Data Portal~\cite{ODPortal} as the \texttt{DoubleMu} primary dataset \cite{opendata}, corresponding to $8.7$~fb$^{-1}$ of integrated luminosity.

As our primary analysis objects, we select the two highest transverse momentum ($p_{\rm T}$) muons that pass the \texttt{MuonTightID} criteria~\cite{CMS:2018rym}.
We also select events that pass the \texttt{HLT\_TrkMu15\_DoubleTrkMu5NoFiltersNoVtx} trigger, which requires at least 2 tracker muons with $p_{\rm T}$'s of at least 15 and 5 GeV, respectively, with no requirement that the muons come from the same primary vertex.
To avoid trigger turn-on effects, we enforce further $p_{\rm T}$ cuts on the muons of 17 and 8 GeV~\cite{CMS:2018jid}.
We refer to the higher (lower) $p_{\rm T}$ muon as the harder (softer) one.
We then divide the data into two channels: \emph{opposite sign} (OS), where the two hardest muons have opposite electric charges (i.e.~$\mu^+\mu^-$), and \emph{same sign} (SS), where they have the same charges (i.e.~$\mu^+\mu^+$ or $\mu^-\mu^-$).
The SS sample will serve as validation samples for \Fig{results_SS}  below.

Many searches for dimuon resonances require the muons to be isolated from the rest of the collision activity.
Since we are interesting in finding $\Upsilon$'s produced through QCD fragmentation, we instead impose an \emph{anti-isolation criterion}.
Within a rapidity-azimuth ($\Delta R$) distance of 0.4 around each of the two muons, we require the ratio of non-muon $p_{\rm T}$ to muon $p_{\rm T}$ to be \emph{greater} than $0.55$.
This threshold was chosen to be as stringent as possible while still retaining enough events to apply ML-based techniques --- approximately 10k events are needed to ensure our models are well trained.
Unlike \Reference{Cooke:2023ukz}, we do not require any matching to jets.
See \Reference{Cesarotti:2019nax} for a previous CMS Open Data study that considered prompt dimuons (i.e.~coming from the collision vertex), which has a substantial anti-isolated component, and \Reference{Witkowski:2023htt} for another CMS Open Data study using ML to study the relation between isolation and promptness.

Anti-isolated $\Upsilon$'s can only be seen at $1.6\sigma$ using the dimuon mass spectrum alone (see the right edge of \Fig{significances_OS} below).
By comparison, without any cuts, $\Upsilon$'s are clearly visible at the $28\sigma$ level.
Anti-isolated $\Upsilon$'s are typically found within QCD jets and are produced promptly during fragmentation.
There are two primary backgrounds to anti-isolated $\Upsilon$'s.
The first is \textit{uncorrelated hadron decays}, where QCD processes produce hadrons (mostly charged pions) that decay in flight to non-prompt muons; this produces approximately equal numbers of SS and OS dimuons.
The second is \textit{Drell-Yan production}, where virtual photons and $Z$ bosons directly yield prompt dimuons via $\gamma/Z^* \to \mu^+ \mu^-$; this produces only OS dimuons and is essentially an irreducible background.
By comparing the OS and SS distributions (\Figs{histogram_SS2}{histogram_OS} below), we estimate that these two backgrounds are roughly the same size in the anti-isolated OS channel.

To enhance the visibility of anti-isolated $\Upsilon$'s, we consider three auxiliary features --- the dimuon transverse momentum ${p_{\rm T}}_{\mu^+\mu^-}$ and the 3D impact parameters (IP3D) of the harder and softer muons --- which we refer to collectively as $x$.
This feature set was chosen both for its performance in reconstructing the $\Upsilon$ and because it does not produce spurious peaks or sculpting artifacts.
Other feature sets we considered include various combinations of single- and di-muon kinematic observables, as well as $\Delta R_{\mu^+\mu^-}$ and $\Delta{p_{\rm T}}_{\mu^+\mu^-}$.
Because of the limited size of the sideband training data, using too many features leads to suboptimal CATHODE performance.


\begin{figure*}[tpbh]
    \centering
    \subfloat[]{
         \includegraphics[width=0.45\textwidth]{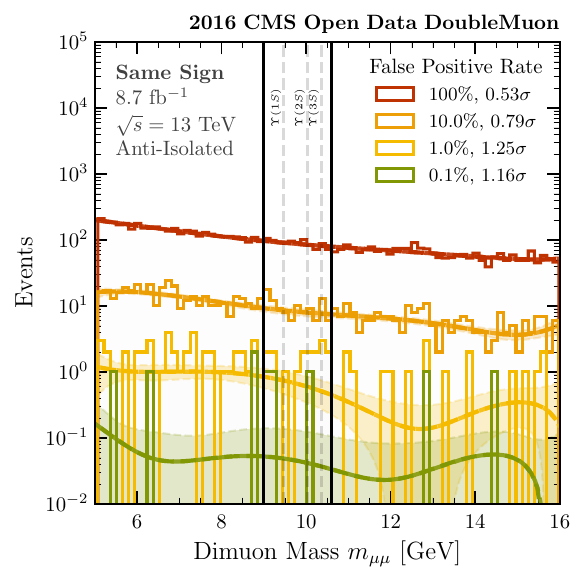}
        \label{fig:histogram_SS2}
    }
    $\qquad$
    \subfloat[]{
        \includegraphics[width=0.45\textwidth]{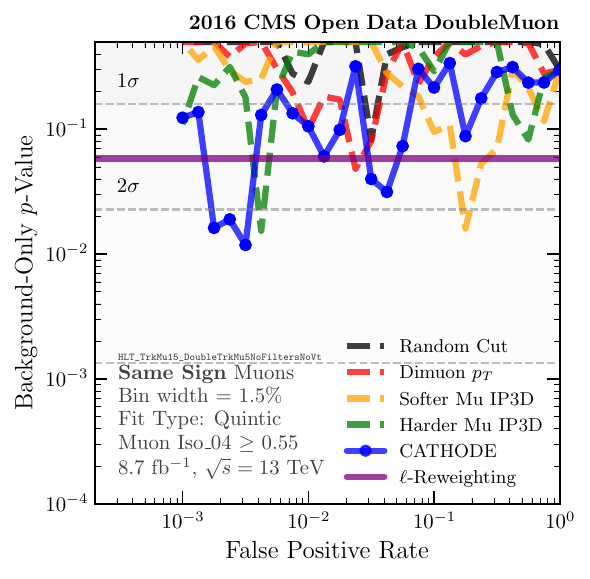}
        \label{fig:significances_SS2}
    }
    \caption{
         Results of the SS validation study.
         (a) Dimuon mass distributions after a series of cuts on the BDT classifier for different FPR working points.
         For each FPR, a quintic polynomial is fit to the two SB regions and interpolated into the SR, delineated by the vertical black lines.
         (b) The $p$-value (significance) as a function of FPR for \Cathode{} (blue), cuts on the dimuon $p_{\rm T}$ (red) and the muon 3D impact parameters (yellow and green), and a random cut as a baseline (black).  
        }
    \label{fig:results_SS}
\end{figure*}

\textbf{\textit{General Analysis Procedure.}}
The core assumptions of our analysis is that the signal of interest is positive,  localized in the dimuon mass $m \equiv m_{\mu+ \mu-}$, and lives atop a smooth background. 
Even though we expect three distinct $\Upsilon$ peaks, they count as one resonant feature for a wide enough \emph{signal region} (SR).
In the sidebands (SBs), we assume the data is predominantly background, and we distinguish the \emph{low-mass sideband} (SBL) from the \emph{high-mass sideband} (SBH).
These regions are separated by vertical black lines in \Figs{histogram_SS2}{histogram_OS} below.

For an ML-based resonance search (first outlined in \cite{Collins:2018epr,Collins_2019}), or \emph{extended bump hunt}, the strategy is to mask the SR and only use the SB information (and the assumption of smoothness) to construct an estimate of the background distribution $p_{\rm Bkg}(x,m)$ for a set of auxiliary features $x$. 
With these assumptions, we can compare the observed data $p_{\rm Data}(x,m)$ to the background-only null hypothesis $p_{\rm Bkg}(x,m)$ in the SR to test for possible resonances.
We compare our ML-based results to ``classical'' cuts on single features, for which only Steps 1 and 4a below are needed.

The full analysis procedure is as follows:
\begin{itemize}
    \item[1.] \textbf{Choose Signal and Sideband Regions}. 
   To define the bump hunt window, we choose the SR to contain the three dimuon $\Upsilon$ resonances and the SBL and SBH to avoid other known dimuon resonances, corresponding to:
    $\text{SBL} = [5.0, 9.0]\,\GeV$, $\text{SR} = [9.0, 10.6]\,\GeV$, and $\text{SBH} = [10.6, 16.0]\,\GeV$.  

    \item[2.] \textbf{Interpolate Background into SR.}
    Following $\Cathode$~\cite{Hallin:2021wme},
    we train an ensemble of Normalizing Flows (NFs)~\cite{cms1266935020, Kobyzev_2021, papamakarios2021normalizingflowsprobabilisticmodeling} on the SBL and SBH data (with the SR masked out) to learn the conditional distribution $p_{\rm Bkg}(x|m)$.
    To estimate $p_{\rm Bkg}(m)$,
    we fit a quintic polynomial to a histogram of $m$ in the SBs, using a relative bin width of 1.5\%.
    This width was chosen to be above the $1\%$ dimuon mass resolution (verified via $J/\psi$ fits and consistent with similar dimuon studies \cite{Cesarotti:2019nax, CMS:2023slr}).
    We can then generate interpolated background events in the SR according to $p_{\rm Bkg}(x, m) \equiv p_{\rm Bkg}(x|m)\, p_{\rm Bkg}(m)$ by sampling the NFs weighted by the polynomial mass fit.

    \item[3.] \textbf{Calculate Likelihood Ratios.} 
    Restricting our focus to the SR, 
    we train an ensemble of Boosted Decision Trees (BDTs)~\cite{BreiFrieStonOlsh84, friedman2000greedy} to 
    distinguish observed data $p_{\rm Data}(x)$ from the interpolated background $p_{\rm Bkg}(x)$.
    To do this, BDTs learn optimal sequences of cuts on their inputs.
    We choose BDTs over other architectures because they are fast to train, and they were found to be especially robust to noise in previous extended bump hunt studies~\cite{Finke:2023ltw,Freytsis:2023cjr}. 
    We do not provide the BDT with mass information.
    We assume the data distribution can be expressed as $p_{\rm Data}(x) = \mu \, p_{\rm Sig}(x) + (1-\mu) \, p_{\rm Bkg}(x)$, where $\mu = \frac{N_{\rm Sig}}{N_{\rm Sig} + N_{\rm Bkg}}$ is the signal fraction.
    In the asymptotic limit, the classifier score $z(x) = \frac{p_{\rm Data}(x)}{p_{\rm Data}(x) + p_{\rm Bkg}(x)}$ is monotonically related to the likelihood ratio:
    \begin{equation}
        \label{eq:ell_def}
        \ell(x) \equiv \frac{z(x) - (1-\mu)\,(1-z(x))}{\mu\,(1-z(x))} = \frac{p_{\rm Sig}(x)}{p_{\rm Bkg}(x)},
    \end{equation}
    
    which in turn is the most powerful mass-independent test statistic~\cite{NP, Metodiev:2017vrx}. 
    To make the most of the limited SR data, we use 5-fold cross-validation~\cite{5360332} to ensure that the classifiers are never trained on a subset of data they will evaluate.
    \item[4.] \textbf{Estimate Significances.} We have two different methods to estimate the signal significance:
    \item[a.] \textbf{Cut and Count.}
    Here, we cut on a feature of interest, count the number of events that pass the cut, and compare to the background estimate.
    We place a cut either on the classifier score for the ML-based approach or on individual features for the classical comparisons.
    Working points are defined by the corresponding False-Positive Rate (FPR) in the SBs.
    To estimate the number of background events in the SR after the cut, we fit the SB mass spectrum using another quintic fit as above.
    Given the number of observed events in the SR after the cut, we estimate the one-sided $p$-value to reject the background-only hypothesis following \Reference{Cowan:2010js}.
    \item[b.] \textbf{Reweight by the Likelihood.}
    Instead of a single FPR working point, we can do a weighted analysis across different cut values.
    Following \Reference{Freytsis:2009bh} in adapting the Matrix Element Method~\cite{Kondo:1988yd, Kondo:1991dw}, we assign each event $i$ a weight $w_i = \ell(x_i)$ related to the estimated signal-to-background likelihood ratio from \Eq{ell_def}.
    These weights generalize the FPR cut in Step 4a: there, events have weight 1 if they pass a given threshold and 0 otherwise.
    Because the estimate of $\ell$ is imperfect, some weights may be negative, corresponding to the most background-like events. 
    We impose a mild classifier cut to remove these negative weight events (equivalently, set negative weights to zero).
    We then fit the mass spectrum of the surviving events in the SB to estimate the background-only $p$-value in the SR --- this is just like the fits of Step 4a, but now with weighted events.
    The Poisson binned likelihoods of \Reference{Cowan:2010js} are replaced with Scaled Poisson Distributions~\cite{Bohm:2013gla} to account for the event weights.
\end{itemize}

The likelihood reweighting in Step 4b is expected to be more powerful than the cut-and-count analysis in Step 4a~\cite{Freytsis:2009bh}.
Step 4b, however, can only be performed for the ML-based method, since obtaining an estimate for the likelihood ratio requires an estimate of $p_{\rm Bkg}(x,m)$, which is only accessible via the NF.
If the classifier-learned $\ell$ is a poor estimate of the likelihood ratio, this method is still valid, albeit with degraded statistical power.

To mitigate issues related to ML training, we take ensemble averages of our models (5 NFs and 100 BDTs).
The NFs are ensembled after Step 2 by drawing samples equally from each NF, and the BDTs are ensembled after training in Step 3 by averaging their scores.
For the quintic polynomial fits, we profile over the fit parameters.
We briefly discuss alternate analysis choices in the End Matter, in particular the mass fit form and bin width.


\begin{figure*}[tpbh]
    \centering
    \subfloat[]{
         \includegraphics[width=0.45\textwidth]{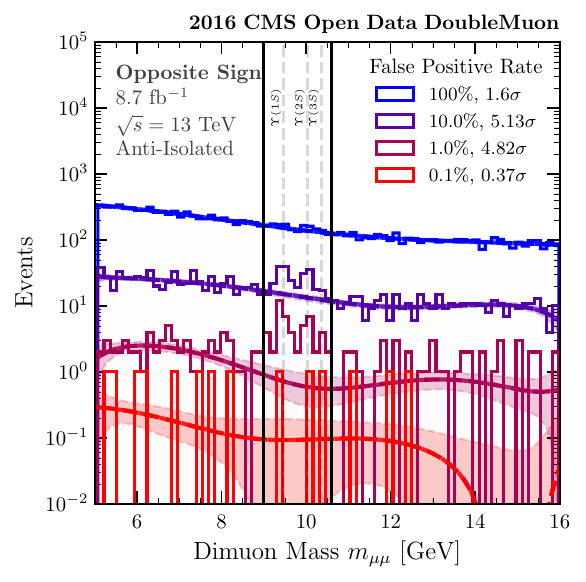}
        \label{fig:histogram_OS}
    }
    $\qquad$
    \subfloat[]{
        \includegraphics[width=0.45\textwidth]{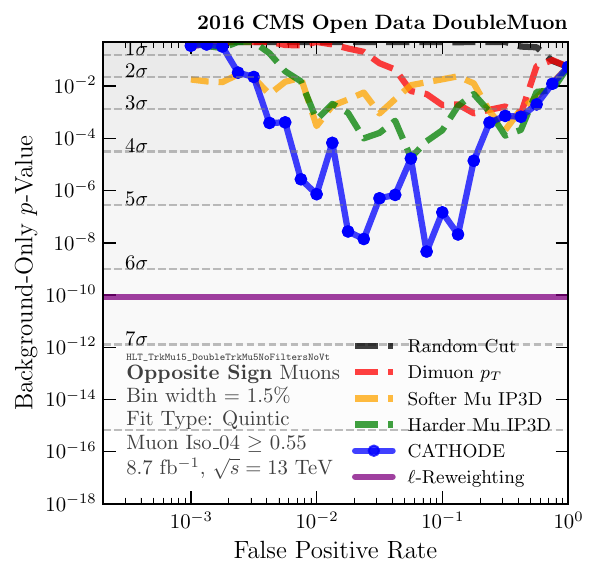}
        \label{fig:significances_OS}
    }
    \caption{
        Results of the OS bump hunt analysis, shown in the same format as \Fig{results_SS}.
        (a) After a cut on the BDT classifier, clear peaks emerge in the SR above the SB-fitted background.
        (b) Noting the different $y$-axis scale from \Fig{significances_SS2}, the $\Upsilon$ significance grows from $1.5\sigma$ to well over $5\sigma$ with CATHODE, especially with likelihood reweighting.  
        }
    \label{fig:results_OS}
\end{figure*}

\textbf{\textit{Same-Sign Validation Study.}}
To be confident in the statistical validity of our methods, we apply our analysis procedure to a ``control'' sample of SS dimuons, with roughly half the number of events as the OS sample.
A doubly-charged resonance decaying to muons ($\Phi^{++} \to \mu^+ \mu^+$) at $\sim 10$ GeV is strongly constrained by precision electroweak tests, such as the $Z$-pole width and running of $\alpha_{\rm EM}$~\cite{ParticleDataGroup:2024cfk}.
Such a resonance is also ruled out by model-specific searches for doubly-charged scalars~\cite{Swartz:1990ki, Atag:2003wk}.
Thus, we seek to verify that our procedure \emph{does not} find a signal in the SS channel where we expect there to be none.

In \Fig{histogram_SS2}, we show the dimuon mass spectrum after various BDT cuts.
As we impose stricter criteria on the FPR (i.e.~smaller fraction of background events passing the cut), no significant peaks are observed in the SR.
We quantify this via the significance curves in \Fig{significances_SS2}.
The cut-and-count significances (Step 4a) are plotted as a function of the FPR, for both CATHODE and the classical tests, and no method finds evidence of a localized excess above 2.4$\sigma$.
The ML-based likelihood reweighted significance (Step 4b) is plotted as a horizontal line at a modest $1.6\sigma$.
We conclude that our method successfully avoids sculpting spurious signals.

For completeness, we perform two additional validations.
First, we use a classifier trained on OS data to look for resonances in SS data. 
We (successfully) do not find a SS signal, which implies that a classifier trained on (signal-containing) OS data does not sculpt peaks where none exist.
Second, we use a classifier trained on SS data to look for resonances in OS data.
There is indeed a initial 1.6$\sigma$ $\Upsilon$ signal in the OS channel (as discussed below), but we (successfully) do not elevate the signal significance.
We conclude that in order to reveal the $\Upsilon$ signal, the BDT classifier must be able to learn nontrivial correlations between auxiliary features, which are not present in the SS control sample.


\begin{figure*}[t]
    \centering
    \includegraphics[width=0.99\linewidth]{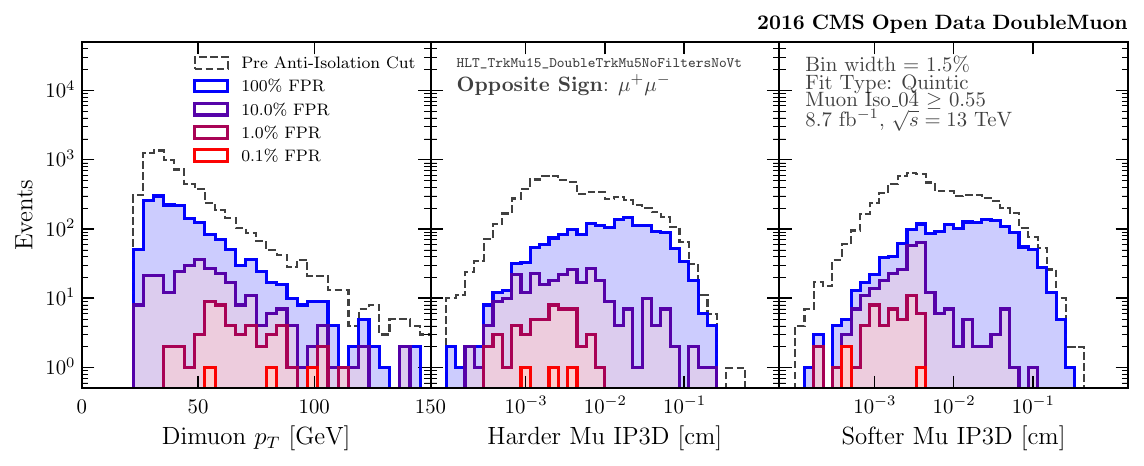}
    \caption{Distributions in the SR of three auxiliary features --- dimuon $p_{\rm T}$, harder muon IP3D, and softer muons IP3D --- after increasingly stringent cuts on the BDT classifier score.
    The dashed curves show  results before the initial anti-isolation cut.}
    \label{fig:features_OS}
\end{figure*}

\textbf{\textit{Opposite-Sign Search Results.}}
We now present search results in the OS channel for anti-isolated $\Upsilon \to \mu^+ \mu^-$ decays.
In \Fig{histogram_OS}, we show the dimuon mass distribution after a sequence of BDT cuts.
A modest initial excess in the SR is visually amplified by cuts on the classifier.
The quantitative gains in signal significance are shown in \Fig{significances_OS}.
\Cathode{} achieves a maximum significance of $5.7\sigma$ at a 7.5\% FPR working point, which is increased to $6.4\sigma$ with likelihood reweighting.
By contrast, none of the classical  cuts surpass the nominal 5$\sigma$ discovery threshold --- cutting on the harder muon's IP3D achieves at most $4.1\sigma$ significance.
This demonstrates substantial gains from using a multidimensional ML-based approach compared to single-feature classical cuts.
See the End Matter for a study of multi-feature classical cuts.


To better understand how the BDT discriminates $\Upsilon$-like events from background, we can inspect the impact of classifier cuts on the auxiliary features in the SR, shown in \Fig{features_OS}.
With more stringent FPR cuts, the BDT selects events with moderate dimuon $p_{\rm T}$ ($\approx 60$ GeV) and small IP3Ds ($\approx 10^{-3}$ cm), similar to the phase space selection in \Reference{Cesarotti:2019nax}.
In essence, the BDT is attempting to undo the effects of the initial anti-isolation condition as best as it can to recover the $\Upsilon$ resonance.
By focusing on small impact parameters, the BDT is mitigating the background from uncorrelated hadron decays.
However, cutting on the IP3D alone does not reproduce the original peak at $10^{-3}$ cm; if it did, a classical IP3D cut would have matched the performance of \Cathode{}.
By focusing on moderate dimuon $p_{\rm T}$, the BDT is additionally rejecting low $p_{\rm T}$ backgrounds.
Both the uncorrelated hadron and Drell-Yan backgrounds fall off at high $p_{\rm T}$, although the uncorrelated hadron background does so more sharply, which we can confirm through studies of the SS control sample.
Since the Drell-Yan background generates genuine prompt $\mu^+\mu^-$ signatures, the BDT would need access to additional kinematic features to mitigate it further.


\textbf{\textit{Conclusions.}}
Using an ML-based anomaly detection technique, we elevated an anti-isolated $\Upsilon \to \mu^+\mu^-$ signal to over $5\sigma$ significance in CMS Open Data. 
Our study represents the first analysis of anti-isolated $\Upsilon$ production at the LHC.
This production channel is particularly useful as a probe for the transition region of QCD, and our work is a step towards measuring observables relevant for bottomonium fragmentation.
Moreover, we showed that the statistical performance of multi-feature ML-based methods is superior to that of classical single-feature cut and count, with the best performance coming from reweighting events according to the estimated signal-to-background likelihood.
This is all done without any simulation of signal or background.
Without these novel methods, anti-isolated $\Upsilon$'s are hard to find: the naive classical cuts we tested yield at best around $4\sigma$ significance.

There are a number of ways we could enhance the statistical sensitivity of our analysis.
We chose fixed SR and SBs motivated by the known location of the $\Upsilon$ and other QCD dimuon resonances, but one could  consider several candidate regions to find the optimal choice. 
When estimating backgrounds and likelihoods using NFs and BDTs, we used ensembling to mitigate ML training artifacts, but one could account for both statistical and systematic uncertainties in the architecture and training.
This would improve sensitivity for the likelihood reweighting method, where there are uncertainties from the estimate of $\mu$ in \Eq{ell_def} lead to suboptimal (though still valid) weights.
While we do not expect these variations to substantially improve the quoted significances, they could be relevant for revealing more elusive signals.

Now that we have ``rediscovered'' the $\Upsilon$, we plan to build upon this study in the future and perform a full scan over the dimuon mass range to search for resonances beyond the Standard Model.
We encourage the anomaly detection community to use this study and this dataset for development and testing, since the $\Upsilon$ peak serves as a ``standard candle'' whose properties are well known.
This dataset is complementary to previously publicized benchmark datasets, since it corresponds to real, noisy, detector-level experimental data.
It also presents new challenges; for example, we found that not all auxiliary feature sets avoided sculpting, a hurdle which we had not encountered in previous studies on synthetic data.
We hope this $\Upsilon$ analysis inspires further anomaly detection studies in real collider data. 

\begin{acknowledgments}
\textit{Acknowledgments.}
We would like to thank Ed Witkowski for early discussions and contributions to this work, and Cari Cesarotti, Dennis Noll, Matt Strassler, and Daniel Whiteson for detailed feedback and encouragement.
RG and JT are supported by the National Science Foundation (NSF) under Cooperative Agreement PHY-2019786 (The NSF AI Institute for Artificial Intelligence and Fundamental Interactions (\url{http://iaifi.org/}), and by the U.S.\ Department of Energy (DOE) Office of High Energy Physics under grant number DE-SC0012567.
JT is additionally supported by the Simons Foundation
through Investigator grant 929241, and in part by grant NSF PHY-2309135 to the Kavli Institute for Theoretical Physics (KITP).
RM and BN are supported by the U.S.~DOE Office of Science under contract DE-AC02-05CH11231 and Grant No.~63038 from the John Templeton Foundation.
This research used resources of the National Energy Research Scientific Computing Center, a DOE Office of Science User Facility supported by the Office of Science of the U.S. Department of Energy under Contract No. DE-AC02-05CH11231 using NERSC award HEP-ERCAP0021099.

\end{acknowledgments}

\textbf{\textit{End Matter: Code and Data.}}
All of the code to reproduce the analyses and figures in this paper can be found at \url{https://github.com/hep-lbdl/dimuonAD}.
The code consists of a series of numbered Python files and Jupyter notebooks, that, when run in order, will completely reproduce our analyses from scratch. 
The BDTs are trained using the \textsc{XGBoost} package \cite{Chen_2016} and the \Cathode{} normalizing flow is built and trained with \textsc{PyTorch} \cite{Ansel_PyTorch_2_Faster_2024}. The architectures and hyperparameters used in this analysis can be found in the \textsc{GitHub} repository in the \texttt{configs} folder.
The datasets used in this paper are available on Zenodo at \url{https://zenodo.org/records/14618719}.

\textbf{\textit{End Matter: Systematic Variations and Pseudoexperiments.}}
To test the robustness of our anomaly detection strategy, we want to show its behavior under systematic variations. 
The baseline results in the main text were based on a quintic polynomial background fit and a dimuon relative bin width of 1.5\%.
For fit variations, we consider cubic and septic polynomials.
For bin width variations, we consider 1.1\% and 2.3\%.

\begin{figure}
    \centering\includegraphics[width=0.99\linewidth]{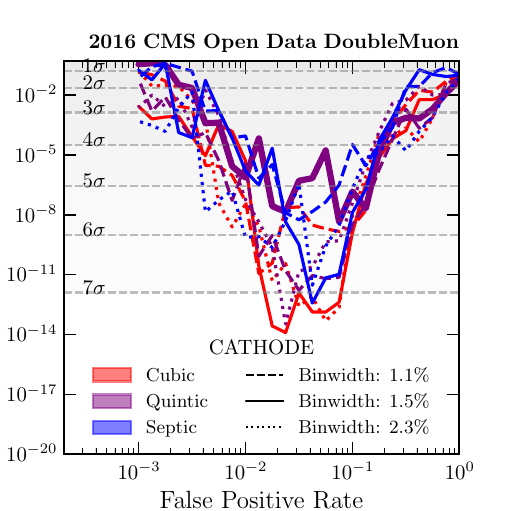}
    \caption{The \Cathode{} $p$-values as a function of FPR, for different choices of fit polynomial and bin width. The central fit and binning choice used in \Fig{significances_OS} is bolded in solid purple.}
    \label{fig:variations}
\end{figure}

In \Fig{variations}, we plot the \Cathode{} $p$-value as a function of FPR for these different variations, analogous to the solid blue line in \Fig{significances_OS}. 
All of these variations yield similar results to the baseline approach, and all choices considered surpass the 5$\sigma$ discovery threshold.
This shows that the $\Upsilon$ significances are robust with respect to these choices.
If anything, the baseline approach is conservative relative to these variations, likely due to a statistical fluctuation.
We note that similar robustness to variations holds for the single-feature classical cuts and for the likelihood reweighting method.

Another potential concern is whether our computed $p$-values are valid even though we are using an ML-based approach.
To verify that the standard asymptotic formulae~\cite{Cowan:2010js} for $p$-values are still valid, we did a numerical analysis with pseudoexperiments.
Specifically, we generated 1000 pseudoexperiments by sampling from a trained NF, which we take to be our null hypothesis.
We then computed the observed distribution of the test statistic, which follows the expected asymptotic distribution out to at least the $3\sigma$ level.

\textbf{\textit{End Matter: Scan Over Multiple Cuts.}}
Given the benefits seen from using multi-dimensional information, it is interesting to check whether classical cuts on multiple features might mimic the CATHODE performance.
Rather than cutting on a single feature, we can scan over simultaneous cuts to see if multi-dimensional auxiliary information can yield a more significant $\Upsilon$ peak. 
We emphasize that this study was carried out after the anomaly detection analysis, so the direction of the cuts was already informed by \Fig{features_OS}.
Nevertheless, it is interesting to see the impact of adding more features without accounting for detailed correlations.

We scan over different cut values for the dimuon $p_{\rm T}$ and the individual IP3D values, where for simplicity, we use the same cut on both muons. 
This is a 2D scan, and we choose 10 points in each dimension for 100 working points in total.
We find a maximum local significance of $5.5\sigma$ with $p_{\rm T} > 45.5$ GeV and IP3D $< 0.1$ cm, though this does not account for the look-elsewhere-effect from considering multiple cut values.
This performance is slightly less than the best CATHODE working point cut (and also less powerful than the likelihood reweighting method), showing that there is a benefit from considering feature correlations.

\bibliography{myrefs,HEPML}

\begin{thebibliography}{48}%
\makeatletter
\providecommand \@ifxundefined [1]{%
 \@ifx{#1\undefined}
}%
\providecommand \@ifnum [1]{%
 \ifnum #1\expandafter \@firstoftwo
 \else \expandafter \@secondoftwo
 \fi
}%
\providecommand \@ifx [1]{%
 \ifx #1\expandafter \@firstoftwo
 \else \expandafter \@secondoftwo
 \fi
}%
\providecommand \natexlab [1]{#1}%
\providecommand \enquote  [1]{``#1''}%
\providecommand \bibnamefont  [1]{#1}%
\providecommand \bibfnamefont [1]{#1}%
\providecommand \citenamefont [1]{#1}%
\providecommand \href@noop [0]{\@secondoftwo}%
\providecommand \href [0]{\begingroup \@sanitize@url \@href}%
\providecommand \@href[1]{\@@startlink{#1}\@@href}%
\providecommand \@@href[1]{\endgroup#1\@@endlink}%
\providecommand \@sanitize@url [0]{\catcode `\\12\catcode `\$12\catcode `\&12\catcode `\#12\catcode `\^12\catcode `\_12\catcode `\%12\relax}%
\providecommand \@@startlink[1]{}%
\providecommand \@@endlink[0]{}%
\providecommand \url  [0]{\begingroup\@sanitize@url \@url }%
\providecommand \@url [1]{\endgroup\@href {#1}{\urlprefix }}%
\providecommand \urlprefix  [0]{URL }%
\providecommand \Eprint [0]{\href }%
\providecommand \doibase [0]{http://dx.doi.org/}%
\providecommand \selectlanguage [0]{\@gobble}%
\providecommand \bibinfo  [0]{\@secondoftwo}%
\providecommand \bibfield  [0]{\@secondoftwo}%
\providecommand \translation [1]{[#1]}%
\providecommand \BibitemOpen [0]{}%
\providecommand \bibitemStop [0]{}%
\providecommand \bibitemNoStop [0]{.\EOS\space}%
\providecommand \EOS [0]{\spacefactor3000\relax}%
\providecommand \BibitemShut  [1]{\csname bibitem#1\endcsname}%
\let\auto@bib@innerbib\@empty
\bibitem [{\citenamefont {Ernstrom}\ and\ \citenamefont {Lonnblad}(1997)}]{Ernstrom:1996am}%
  \BibitemOpen
  \bibfield  {author} {\bibinfo {author} {\bibfnamefont {Per}\ \bibnamefont {Ernstrom}}\ and\ \bibinfo {author} {\bibfnamefont {Leif}\ \bibnamefont {Lonnblad}},\ }\bibfield  {title} {\enquote {\bibinfo {title} {{Generating heavy quarkonia in a perturbative QCD cascade}},}\ }\href {\doibase 10.1007/s002880050446} {\bibfield  {journal} {\bibinfo  {journal} {Z. Phys. C}\ }\textbf {\bibinfo {volume} {75}},\ \bibinfo {pages} {51--56} (\bibinfo {year} {1997})},\ \Eprint {http://arxiv.org/abs/hep-ph/9606472} {arXiv:hep-ph/9606472} \BibitemShut {NoStop}%
\bibitem [{\citenamefont {Andronic}\ \emph {et~al.}(2016)\citenamefont {Andronic} \emph {et~al.}}]{Andronic:2015wma}%
  \BibitemOpen
  \bibfield  {author} {\bibinfo {author} {\bibfnamefont {A.}~\bibnamefont {Andronic}} \emph {et~al.},\ }\bibfield  {title} {\enquote {\bibinfo {title} {{Heavy-flavour and quarkonium production in the LHC era: from proton\textendash{}proton to heavy-ion collisions}},}\ }\href {\doibase 10.1140/epjc/s10052-015-3819-5} {\bibfield  {journal} {\bibinfo  {journal} {Eur. Phys. J. C}\ }\textbf {\bibinfo {volume} {76}},\ \bibinfo {pages} {107} (\bibinfo {year} {2016})},\ \Eprint {http://arxiv.org/abs/1506.03981} {arXiv:1506.03981 [nucl-ex]} \BibitemShut {NoStop}%
\bibitem [{\citenamefont {Bain}\ \emph {et~al.}(2016)\citenamefont {Bain}, \citenamefont {Dai}, \citenamefont {Hornig}, \citenamefont {Leibovich}, \citenamefont {Makris},\ and\ \citenamefont {Mehen}}]{Bain:2016clc}%
  \BibitemOpen
  \bibfield  {author} {\bibinfo {author} {\bibfnamefont {Reggie}\ \bibnamefont {Bain}}, \bibinfo {author} {\bibfnamefont {Lin}\ \bibnamefont {Dai}}, \bibinfo {author} {\bibfnamefont {Andrew}\ \bibnamefont {Hornig}}, \bibinfo {author} {\bibfnamefont {Adam~K.}\ \bibnamefont {Leibovich}}, \bibinfo {author} {\bibfnamefont {Yiannis}\ \bibnamefont {Makris}}, \ and\ \bibinfo {author} {\bibfnamefont {Thomas}\ \bibnamefont {Mehen}},\ }\bibfield  {title} {\enquote {\bibinfo {title} {{Analytic and Monte Carlo Studies of Jets with Heavy Mesons and Quarkonia}},}\ }\href {\doibase 10.1007/JHEP06(2016)121} {\bibfield  {journal} {\bibinfo  {journal} {JHEP}\ }\textbf {\bibinfo {volume} {06}},\ \bibinfo {pages} {121} (\bibinfo {year} {2016})},\ \Eprint {http://arxiv.org/abs/1603.06981} {arXiv:1603.06981 [hep-ph]} \BibitemShut {NoStop}%
\bibitem [{\citenamefont {Celiberto}\ and\ \citenamefont {Fucilla}(2022)}]{Celiberto:2022dyf}%
  \BibitemOpen
  \bibfield  {author} {\bibinfo {author} {\bibfnamefont {Francesco~Giovanni}\ \bibnamefont {Celiberto}}\ and\ \bibinfo {author} {\bibfnamefont {Michael}\ \bibnamefont {Fucilla}},\ }\bibfield  {title} {\enquote {\bibinfo {title} {{Diffractive semi-hard production of a $J/\psi $ or a $\Upsilon $ from single-parton fragmentation plus a jet in hybrid factorization}},}\ }\href {\doibase 10.1140/epjc/s10052-022-10818-8} {\bibfield  {journal} {\bibinfo  {journal} {Eur. Phys. J. C}\ }\textbf {\bibinfo {volume} {82}},\ \bibinfo {pages} {929} (\bibinfo {year} {2022})},\ \Eprint {http://arxiv.org/abs/2202.12227} {arXiv:2202.12227 [hep-ph]} \BibitemShut {NoStop}%
\bibitem [{\citenamefont {Aaij}\ \emph {et~al.}(2017)\citenamefont {Aaij} \emph {et~al.}}]{LHCb:2017llq}%
  \BibitemOpen
  \bibfield  {author} {\bibinfo {author} {\bibfnamefont {Roel}\ \bibnamefont {Aaij}} \emph {et~al.} (\bibinfo {collaboration} {LHCb}),\ }\bibfield  {title} {\enquote {\bibinfo {title} {{Study of J/\ensuremath{\psi} Production in Jets}},}\ }\href {\doibase 10.1103/PhysRevLett.118.192001} {\bibfield  {journal} {\bibinfo  {journal} {Phys. Rev. Lett.}\ }\textbf {\bibinfo {volume} {118}},\ \bibinfo {pages} {192001} (\bibinfo {year} {2017})},\ \Eprint {http://arxiv.org/abs/1701.05116} {arXiv:1701.05116 [hep-ex]} \BibitemShut {NoStop}%
\bibitem [{\citenamefont {Tumasyan}\ \emph {et~al.}(2022)\citenamefont {Tumasyan} \emph {et~al.}}]{CMS:2021puf}%
  \BibitemOpen
  \bibfield  {author} {\bibinfo {author} {\bibfnamefont {Armen}\ \bibnamefont {Tumasyan}} \emph {et~al.} (\bibinfo {collaboration} {CMS}),\ }\bibfield  {title} {\enquote {\bibinfo {title} {{Fragmentation of jets containing a prompt J$/\psi$ meson in PbPb and pp collisions at $\sqrt{s_\mathrm{NN}} =$ 5.02 TeV}},}\ }\href {\doibase 10.1016/j.physletb.2021.136842} {\bibfield  {journal} {\bibinfo  {journal} {Phys. Lett. B}\ }\textbf {\bibinfo {volume} {825}},\ \bibinfo {pages} {136842} (\bibinfo {year} {2022})},\ \Eprint {http://arxiv.org/abs/2106.13235} {arXiv:2106.13235 [hep-ex]} \BibitemShut {NoStop}%
\bibitem [{\citenamefont {Bain}\ \emph {et~al.}(2017)\citenamefont {Bain}, \citenamefont {Dai}, \citenamefont {Leibovich}, \citenamefont {Makris},\ and\ \citenamefont {Mehen}}]{Bain:2017wvk}%
  \BibitemOpen
  \bibfield  {author} {\bibinfo {author} {\bibfnamefont {Reggie}\ \bibnamefont {Bain}}, \bibinfo {author} {\bibfnamefont {Lin}\ \bibnamefont {Dai}}, \bibinfo {author} {\bibfnamefont {Adam}\ \bibnamefont {Leibovich}}, \bibinfo {author} {\bibfnamefont {Yiannis}\ \bibnamefont {Makris}}, \ and\ \bibinfo {author} {\bibfnamefont {Thomas}\ \bibnamefont {Mehen}},\ }\bibfield  {title} {\enquote {\bibinfo {title} {{NRQCD Confronts LHCb Data on Quarkonium Production within Jets}},}\ }\href {\doibase 10.1103/PhysRevLett.119.032002} {\bibfield  {journal} {\bibinfo  {journal} {Phys. Rev. Lett.}\ }\textbf {\bibinfo {volume} {119}},\ \bibinfo {pages} {032002} (\bibinfo {year} {2017})},\ \Eprint {http://arxiv.org/abs/1702.05525} {arXiv:1702.05525 [hep-ph]} \BibitemShut {NoStop}%
\bibitem [{\citenamefont {Bodwin}\ \emph {et~al.}(1995)\citenamefont {Bodwin}, \citenamefont {Braaten},\ and\ \citenamefont {Lepage}}]{Bodwin:1994jh}%
  \BibitemOpen
  \bibfield  {author} {\bibinfo {author} {\bibfnamefont {Geoffrey~T.}\ \bibnamefont {Bodwin}}, \bibinfo {author} {\bibfnamefont {Eric}\ \bibnamefont {Braaten}}, \ and\ \bibinfo {author} {\bibfnamefont {G.~Peter}\ \bibnamefont {Lepage}},\ }\bibfield  {title} {\enquote {\bibinfo {title} {{Rigorous QCD analysis of inclusive annihilation and production of heavy quarkonium}},}\ }\href {\doibase 10.1103/PhysRevD.55.5853} {\bibfield  {journal} {\bibinfo  {journal} {Phys. Rev. D}\ }\textbf {\bibinfo {volume} {51}},\ \bibinfo {pages} {1125--1171} (\bibinfo {year} {1995})},\ \bibinfo {note} {[Erratum: Phys.Rev.D 55, 5853 (1997)]},\ \Eprint {http://arxiv.org/abs/hep-ph/9407339} {arXiv:hep-ph/9407339} \BibitemShut {NoStop}%
\bibitem [{\citenamefont {Cooke}(2023)}]{Cooke:2023ukz}%
  \BibitemOpen
  \bibfield  {author} {\bibinfo {author} {\bibfnamefont {Naomi}\ \bibnamefont {Cooke}},\ }\emph {\bibinfo {title} {Measurements of Quarkonia and Tetraquark Production in Jets at LHCb}},\ \href@noop {} {Ph.D. thesis},\ \bibinfo  {school} {Glasgow U.} (\bibinfo {year} {2023})\BibitemShut {NoStop}%
\bibitem [{ope()}]{opendata}%
  \BibitemOpen
  \href@noop {} {\ }\bibinfo {note} {CMS Collaboration (2024). DoubleMuon primary dataset in NANOAOD format from RunH of 2016 (/DoubleMuon/Run2016H-UL2016\_MiniAODv2\_NanoAODv9-v1/NANOAOD). CERN Open Data Portal. DOI:10.7483/OPENDATA.CMS.UZD7.Z50M}\BibitemShut {NoStop}%
\bibitem [{\citenamefont {Hallin}\ \emph {et~al.}(2021)\citenamefont {Hallin}, \citenamefont {Isaacson}, \citenamefont {Kasieczka}, \citenamefont {Krause}, \citenamefont {Nachman}, \citenamefont {Quadfasel}, \citenamefont {Schlaffer}, \citenamefont {Shih},\ and\ \citenamefont {Sommerhalder}}]{Hallin:2021wme}%
  \BibitemOpen
  \bibfield  {author} {\bibinfo {author} {\bibfnamefont {Anna}\ \bibnamefont {Hallin}}, \bibinfo {author} {\bibfnamefont {Joshua}\ \bibnamefont {Isaacson}}, \bibinfo {author} {\bibfnamefont {Gregor}\ \bibnamefont {Kasieczka}}, \bibinfo {author} {\bibfnamefont {Claudius}\ \bibnamefont {Krause}}, \bibinfo {author} {\bibfnamefont {Benjamin}\ \bibnamefont {Nachman}}, \bibinfo {author} {\bibfnamefont {Tobias}\ \bibnamefont {Quadfasel}}, \bibinfo {author} {\bibfnamefont {Matthias}\ \bibnamefont {Schlaffer}}, \bibinfo {author} {\bibfnamefont {David}\ \bibnamefont {Shih}}, \ and\ \bibinfo {author} {\bibfnamefont {Manuel}\ \bibnamefont {Sommerhalder}},\ }\bibfield  {title} {\enquote {\bibinfo {title} {{Classifying Anomalies THrough Outer Density Estimation (CATHODE)}},}\ }\href@noop {} {\  (\bibinfo {year} {2021})},\ \Eprint {http://arxiv.org/abs/2109.00546} {arXiv:2109.00546 [hep-ph]} \BibitemShut {NoStop}%
\bibitem [{\citenamefont {Kasieczka}\ \emph {et~al.}(2021)\citenamefont {Kasieczka} \emph {et~al.}}]{Kasieczka:2021xcg}%
  \BibitemOpen
  \bibfield  {author} {\bibinfo {author} {\bibfnamefont {Gregor}\ \bibnamefont {Kasieczka}} \emph {et~al.},\ }\bibfield  {title} {\enquote {\bibinfo {title} {{The LHC Olympics 2020: A Community Challenge for Anomaly Detection in High Energy Physics}},}\ }\href@noop {} {\  (\bibinfo {year} {2021})},\ \Eprint {http://arxiv.org/abs/2101.08320} {arXiv:2101.08320 [hep-ph]} \BibitemShut {NoStop}%
\bibitem [{\citenamefont {Aarrestad}\ \emph {et~al.}(2021)\citenamefont {Aarrestad} \emph {et~al.}}]{Aarrestad:2021oeb}%
  \BibitemOpen
  \bibfield  {author} {\bibinfo {author} {\bibfnamefont {T.}~\bibnamefont {Aarrestad}} \emph {et~al.},\ }\bibfield  {title} {\enquote {\bibinfo {title} {{The Dark Machines Anomaly Score Challenge: Benchmark Data and Model Independent Event Classification for the Large Hadron Collider}},}\ }\href@noop {} {\  (\bibinfo {year} {2021})},\ \Eprint {http://arxiv.org/abs/2105.14027} {arXiv:2105.14027 [hep-ph]} \BibitemShut {NoStop}%
\bibitem [{liv()}]{livingReview}%
  \BibitemOpen
  \href@noop {} {\ }\bibinfo {note} {Living Review of Machine Learning in High Energy Physics. \url{https://iml-wg.github.io/HEPML-LivingReview}}\BibitemShut {NoStop}%
\bibitem [{\citenamefont {Aad}\ \emph {et~al.}(2020)\citenamefont {Aad} \emph {et~al.}}]{ATLAS:2020iwa}%
  \BibitemOpen
  \bibfield  {author} {\bibinfo {author} {\bibfnamefont {Georges}\ \bibnamefont {Aad}} \emph {et~al.} (\bibinfo {collaboration} {ATLAS}),\ }\bibfield  {title} {\enquote {\bibinfo {title} {{Dijet resonance search with weak supervision using $\sqrt{s}=13$ TeV $pp$ collisions in the ATLAS detector}},}\ }\href {\doibase 10.1103/PhysRevLett.125.131801} {\bibfield  {journal} {\bibinfo  {journal} {Phys. Rev. Lett.}\ }\textbf {\bibinfo {volume} {125}},\ \bibinfo {pages} {131801} (\bibinfo {year} {2020})},\ \Eprint {http://arxiv.org/abs/2005.02983} {arXiv:2005.02983 [hep-ex]} \BibitemShut {NoStop}%
\bibitem [{\citenamefont {Aad}\ \emph {et~al.}(2023)\citenamefont {Aad} \emph {et~al.}}]{ATLAS:2023azi}%
  \BibitemOpen
  \bibfield  {author} {\bibinfo {author} {\bibfnamefont {Georges}\ \bibnamefont {Aad}} \emph {et~al.} (\bibinfo {collaboration} {ATLAS}),\ }\bibfield  {title} {\enquote {\bibinfo {title} {{Anomaly detection search for new resonances decaying into a Higgs boson and a generic new particle $X$ in hadronic final states using $\sqrt{s} = 13$ TeV $pp$ collisions with the ATLAS detector}},}\ }\href {\doibase 10.1103/PhysRevD.108.052009} {\bibfield  {journal} {\bibinfo  {journal} {Phys. Rev. D}\ }\textbf {\bibinfo {volume} {108}},\ \bibinfo {pages} {052009} (\bibinfo {year} {2023})},\ \Eprint {http://arxiv.org/abs/2306.03637} {arXiv:2306.03637 [hep-ex]} \BibitemShut {NoStop}%
\bibitem [{\citenamefont {Chekhovsky}\ \emph {et~al.}(2024)\citenamefont {Chekhovsky} \emph {et~al.}}]{CMS:2024nsz}%
  \BibitemOpen
  \bibfield  {author} {\bibinfo {author} {\bibfnamefont {Vladimir}\ \bibnamefont {Chekhovsky}} \emph {et~al.} (\bibinfo {collaboration} {CMS}),\ }\bibfield  {title} {\enquote {\bibinfo {title} {{Model-agnostic search for dijet resonances with anomalous jet substructure in proton-proton collisions at $\sqrt{s}$ = 13 TeV}},}\ }\href@noop {} {\  (\bibinfo {year} {2024})},\ \Eprint {http://arxiv.org/abs/2412.03747} {arXiv:2412.03747 [hep-ex]} \BibitemShut {NoStop}%
\bibitem [{\citenamefont {Aad}\ \emph {et~al.}(2024)\citenamefont {Aad} \emph {et~al.}}]{ATLAS:2023ixc}%
  \BibitemOpen
  \bibfield  {author} {\bibinfo {author} {\bibfnamefont {Georges}\ \bibnamefont {Aad}} \emph {et~al.} (\bibinfo {collaboration} {ATLAS}),\ }\bibfield  {title} {\enquote {\bibinfo {title} {{Search for New Phenomena in Two-Body Invariant Mass Distributions Using Unsupervised Machine Learning for Anomaly Detection at s=13\,\,TeV with the ATLAS Detector}},}\ }\href {\doibase 10.1103/PhysRevLett.132.081801} {\bibfield  {journal} {\bibinfo  {journal} {Phys. Rev. Lett.}\ }\textbf {\bibinfo {volume} {132}},\ \bibinfo {pages} {081801} (\bibinfo {year} {2024})},\ \Eprint {http://arxiv.org/abs/2307.01612} {arXiv:2307.01612 [hep-ex]} \BibitemShut {NoStop}%
\bibitem [{\citenamefont {Aad}\ \emph {et~al.}(2025)\citenamefont {Aad} \emph {et~al.}}]{ATLAS:2025obc}%
  \BibitemOpen
  \bibfield  {author} {\bibinfo {author} {\bibfnamefont {Georges}\ \bibnamefont {Aad}} \emph {et~al.} (\bibinfo {collaboration} {ATLAS}),\ }\bibfield  {title} {\enquote {\bibinfo {title} {{Weakly supervised anomaly detection for resonant new physics in the dijet final state using proton-proton collisions at $\sqrt{s}=13$ TeV with the ATLAS detector}},}\ }\href@noop {} {\  (\bibinfo {year} {2025})},\ \Eprint {http://arxiv.org/abs/2502.09770} {arXiv:2502.09770 [hep-ex]} \BibitemShut {NoStop}%
\bibitem [{\citenamefont {Knapp}\ \emph {et~al.}(2020)\citenamefont {Knapp}, \citenamefont {Dissertori}, \citenamefont {Cerri}, \citenamefont {Nguyen}, \citenamefont {Vlimant},\ and\ \citenamefont {Pierini}}]{knapp2020adversarially}%
  \BibitemOpen
  \bibfield  {author} {\bibinfo {author} {\bibfnamefont {Oliver}\ \bibnamefont {Knapp}}, \bibinfo {author} {\bibfnamefont {Guenther}\ \bibnamefont {Dissertori}}, \bibinfo {author} {\bibfnamefont {Olmo}\ \bibnamefont {Cerri}}, \bibinfo {author} {\bibfnamefont {Thong~Q.}\ \bibnamefont {Nguyen}}, \bibinfo {author} {\bibfnamefont {Jean-Roch}\ \bibnamefont {Vlimant}}, \ and\ \bibinfo {author} {\bibfnamefont {Maurizio}\ \bibnamefont {Pierini}},\ }\bibfield  {title} {\enquote {\bibinfo {title} {{Adversarially Learned Anomaly Detection on CMS Open Data: re-discovering the top quark}},}\ }\href {\doibase 10.1140/epjp/s13360-021-01109-4} {\  (\bibinfo {year} {2020}),\ 10.1140/epjp/s13360-021-01109-4},\ \Eprint {http://arxiv.org/abs/2005.01598} {arXiv:2005.01598 [hep-ex]} \BibitemShut {NoStop}%
\bibitem [{\citenamefont {Navas}\ \emph {et~al.}(2024)\citenamefont {Navas} \emph {et~al.}}]{ParticleDataGroup:2024cfk}%
  \BibitemOpen
  \bibfield  {author} {\bibinfo {author} {\bibfnamefont {S.}~\bibnamefont {Navas}} \emph {et~al.} (\bibinfo {collaboration} {Particle Data Group}),\ }\bibfield  {title} {\enquote {\bibinfo {title} {{Review of particle physics}},}\ }\href {\doibase 10.1103/PhysRevD.110.030001} {\bibfield  {journal} {\bibinfo  {journal} {Phys. Rev. D}\ }\textbf {\bibinfo {volume} {110}},\ \bibinfo {pages} {030001} (\bibinfo {year} {2024})}\BibitemShut {NoStop}%
\bibitem [{ODP()}]{ODPortal}%
  \BibitemOpen
  \href@noop {} {\ }\bibinfo {note} {CERN Open Data Portal. \url{https://opendata.cern.ch}}\BibitemShut {NoStop}%
\bibitem [{\citenamefont {Sirunyan}\ \emph {et~al.}(2018)\citenamefont {Sirunyan} \emph {et~al.}}]{CMS:2018rym}%
  \BibitemOpen
  \bibfield  {author} {\bibinfo {author} {\bibfnamefont {A.~M.}\ \bibnamefont {Sirunyan}} \emph {et~al.} (\bibinfo {collaboration} {CMS}),\ }\bibfield  {title} {\enquote {\bibinfo {title} {{Performance of the CMS muon detector and muon reconstruction with proton-proton collisions at $\sqrt{s}=$ 13 TeV}},}\ }\href {\doibase 10.1088/1748-0221/13/06/P06015} {\bibfield  {journal} {\bibinfo  {journal} {JINST}\ }\textbf {\bibinfo {volume} {13}},\ \bibinfo {pages} {P06015} (\bibinfo {year} {2018})},\ \Eprint {http://arxiv.org/abs/1804.04528} {arXiv:1804.04528 [physics.ins-det]} \BibitemShut {NoStop}%
\bibitem [{\citenamefont {Sirunyan}\ \emph {et~al.}(2019)\citenamefont {Sirunyan} \emph {et~al.}}]{CMS:2018jid}%
  \BibitemOpen
  \bibfield  {author} {\bibinfo {author} {\bibfnamefont {Albert~M}\ \bibnamefont {Sirunyan}} \emph {et~al.} (\bibinfo {collaboration} {CMS}),\ }\bibfield  {title} {\enquote {\bibinfo {title} {{A search for pair production of new light bosons decaying into muons in proton-proton collisions at 13 TeV}},}\ }\href {\doibase 10.1016/j.physletb.2019.07.013} {\bibfield  {journal} {\bibinfo  {journal} {Phys. Lett. B}\ }\textbf {\bibinfo {volume} {796}},\ \bibinfo {pages} {131--154} (\bibinfo {year} {2019})},\ \Eprint {http://arxiv.org/abs/1812.00380} {arXiv:1812.00380 [hep-ex]} \BibitemShut {NoStop}%
\bibitem [{\citenamefont {Cesarotti}\ \emph {et~al.}(2019)\citenamefont {Cesarotti}, \citenamefont {Soreq}, \citenamefont {Strassler}, \citenamefont {Thaler},\ and\ \citenamefont {Xue}}]{Cesarotti:2019nax}%
  \BibitemOpen
  \bibfield  {author} {\bibinfo {author} {\bibfnamefont {Cari}\ \bibnamefont {Cesarotti}}, \bibinfo {author} {\bibfnamefont {Yotam}\ \bibnamefont {Soreq}}, \bibinfo {author} {\bibfnamefont {Matthew~J.}\ \bibnamefont {Strassler}}, \bibinfo {author} {\bibfnamefont {Jesse}\ \bibnamefont {Thaler}}, \ and\ \bibinfo {author} {\bibfnamefont {Wei}\ \bibnamefont {Xue}},\ }\bibfield  {title} {\enquote {\bibinfo {title} {{Searching in CMS Open Data for Dimuon Resonances with Substantial Transverse Momentum}},}\ }\href {\doibase 10.1103/PhysRevD.100.015021} {\bibfield  {journal} {\bibinfo  {journal} {Phys. Rev. D}\ }\textbf {\bibinfo {volume} {100}},\ \bibinfo {pages} {015021} (\bibinfo {year} {2019})},\ \Eprint {http://arxiv.org/abs/1902.04222} {arXiv:1902.04222 [hep-ph]} \BibitemShut {NoStop}%
\bibitem [{\citenamefont {Witkowski}\ \emph {et~al.}(2023)\citenamefont {Witkowski}, \citenamefont {Nachman},\ and\ \citenamefont {Whiteson}}]{Witkowski:2023htt}%
  \BibitemOpen
  \bibfield  {author} {\bibinfo {author} {\bibfnamefont {Edmund}\ \bibnamefont {Witkowski}}, \bibinfo {author} {\bibfnamefont {Benjamin}\ \bibnamefont {Nachman}}, \ and\ \bibinfo {author} {\bibfnamefont {Daniel}\ \bibnamefont {Whiteson}},\ }\bibfield  {title} {\enquote {\bibinfo {title} {{Learning to isolate muons in data}},}\ }\href {\doibase 10.1103/PhysRevD.108.092008} {\bibfield  {journal} {\bibinfo  {journal} {Phys. Rev. D}\ }\textbf {\bibinfo {volume} {108}},\ \bibinfo {pages} {092008} (\bibinfo {year} {2023})},\ \Eprint {http://arxiv.org/abs/2306.15737} {arXiv:2306.15737 [hep-ex]} \BibitemShut {NoStop}%
\bibitem [{\citenamefont {Collins}\ \emph {et~al.}(2018)\citenamefont {Collins}, \citenamefont {Howe},\ and\ \citenamefont {Nachman}}]{Collins:2018epr}%
  \BibitemOpen
  \bibfield  {author} {\bibinfo {author} {\bibfnamefont {Jack~H.}\ \bibnamefont {Collins}}, \bibinfo {author} {\bibfnamefont {Kiel}\ \bibnamefont {Howe}}, \ and\ \bibinfo {author} {\bibfnamefont {Benjamin}\ \bibnamefont {Nachman}},\ }\bibfield  {title} {\enquote {\bibinfo {title} {{Anomaly Detection for Resonant New Physics with Machine Learning}},}\ }\href {\doibase 10.1103/PhysRevLett.121.241803} {\bibfield  {journal} {\bibinfo  {journal} {Phys. Rev. Lett.}\ }\textbf {\bibinfo {volume} {121}},\ \bibinfo {pages} {241803} (\bibinfo {year} {2018})},\ \Eprint {http://arxiv.org/abs/1805.02664} {arXiv:1805.02664 [hep-ph]} \BibitemShut {NoStop}%
\bibitem [{\citenamefont {Collins}\ \emph {et~al.}(2019)\citenamefont {Collins}, \citenamefont {Howe},\ and\ \citenamefont {Nachman}}]{Collins_2019}%
  \BibitemOpen
  \bibfield  {author} {\bibinfo {author} {\bibfnamefont {Jack~H.}\ \bibnamefont {Collins}}, \bibinfo {author} {\bibfnamefont {Kiel}\ \bibnamefont {Howe}}, \ and\ \bibinfo {author} {\bibfnamefont {Benjamin}\ \bibnamefont {Nachman}},\ }\bibfield  {title} {\enquote {\bibinfo {title} {Extending the search for new resonances with machine learning},}\ }\href {\doibase 10.1103/physrevd.99.014038} {\bibfield  {journal} {\bibinfo  {journal} {Physical Review D}\ }\textbf {\bibinfo {volume} {99}} (\bibinfo {year} {2019}),\ 10.1103/physrevd.99.014038}\BibitemShut {NoStop}%
\bibitem [{\citenamefont {Tabak}\ and\ \citenamefont {Vanden-Eijnden}(2010)}]{cms1266935020}%
  \BibitemOpen
  \bibfield  {author} {\bibinfo {author} {\bibfnamefont {Esteban~G.}\ \bibnamefont {Tabak}}\ and\ \bibinfo {author} {\bibfnamefont {Eric}\ \bibnamefont {Vanden-Eijnden}},\ }\bibfield  {title} {\enquote {\bibinfo {title} {{Density estimation by dual ascent of the log-likelihood}},}\ }\href@noop {} {\bibfield  {journal} {\bibinfo  {journal} {Communications in Mathematical Sciences}\ }\textbf {\bibinfo {volume} {8}},\ \bibinfo {pages} {217 -- 233} (\bibinfo {year} {2010})}\BibitemShut {NoStop}%
\bibitem [{\citenamefont {Kobyzev}\ \emph {et~al.}(2021)\citenamefont {Kobyzev}, \citenamefont {Prince},\ and\ \citenamefont {Brubaker}}]{Kobyzev_2021}%
  \BibitemOpen
  \bibfield  {author} {\bibinfo {author} {\bibfnamefont {Ivan}\ \bibnamefont {Kobyzev}}, \bibinfo {author} {\bibfnamefont {Simon~J.D.}\ \bibnamefont {Prince}}, \ and\ \bibinfo {author} {\bibfnamefont {Marcus~A.}\ \bibnamefont {Brubaker}},\ }\bibfield  {title} {\enquote {\bibinfo {title} {Normalizing flows: An introduction and review of current methods},}\ }\href {\doibase 10.1109/tpami.2020.2992934} {\bibfield  {journal} {\bibinfo  {journal} {IEEE Transactions on Pattern Analysis and Machine Intelligence}\ }\textbf {\bibinfo {volume} {43}},\ \bibinfo {pages} {3964–3979} (\bibinfo {year} {2021})}\BibitemShut {NoStop}%
\bibitem [{\citenamefont {Papamakarios}\ \emph {et~al.}(2021)\citenamefont {Papamakarios}, \citenamefont {Nalisnick}, \citenamefont {Rezende}, \citenamefont {Mohamed},\ and\ \citenamefont {Lakshminarayanan}}]{papamakarios2021normalizingflowsprobabilisticmodeling}%
  \BibitemOpen
  \bibfield  {author} {\bibinfo {author} {\bibfnamefont {George}\ \bibnamefont {Papamakarios}}, \bibinfo {author} {\bibfnamefont {Eric}\ \bibnamefont {Nalisnick}}, \bibinfo {author} {\bibfnamefont {Danilo~Jimenez}\ \bibnamefont {Rezende}}, \bibinfo {author} {\bibfnamefont {Shakir}\ \bibnamefont {Mohamed}}, \ and\ \bibinfo {author} {\bibfnamefont {Balaji}\ \bibnamefont {Lakshminarayanan}},\ }\href {https://arxiv.org/abs/1912.02762} {\enquote {\bibinfo {title} {Normalizing flows for probabilistic modeling and inference},}\ } (\bibinfo {year} {2021}),\ \Eprint {http://arxiv.org/abs/1912.02762} {arXiv:1912.02762 [stat.ML]} \BibitemShut {NoStop}%
\bibitem [{CMS(2023)}]{CMS:2023slr}%
  \BibitemOpen
  \bibfield  {title} {\enquote {\bibinfo {title} {{Search for prompt production of a GeV scale resonance decaying to a pair of muons in proton-proton collisions at $\sqrt{s}=13$ TeV}},}\ }\href@noop {} {\bibfield  {journal} {\bibinfo  {journal} {CMS-PAS-EXO-21-005}\ } (\bibinfo {year} {2023})}\BibitemShut {NoStop}%
\bibitem [{\citenamefont {Breiman}\ \emph {et~al.}(1984)\citenamefont {Breiman}, \citenamefont {Friedman}, \citenamefont {Stone},\ and\ \citenamefont {Olshen}}]{BreiFrieStonOlsh84}%
  \BibitemOpen
  \bibfield  {author} {\bibinfo {author} {\bibfnamefont {Leo}\ \bibnamefont {Breiman}}, \bibinfo {author} {\bibfnamefont {Jerome}\ \bibnamefont {Friedman}}, \bibinfo {author} {\bibfnamefont {Charles~J.}\ \bibnamefont {Stone}}, \ and\ \bibinfo {author} {\bibfnamefont {R.A.}\ \bibnamefont {Olshen}},\ }\href@noop {} {\emph {\bibinfo {title} {Classification and Regression Trees}}}\ (\bibinfo  {publisher} {Chapman and Hall/CRC},\ \bibinfo {year} {1984})\BibitemShut {NoStop}%
\bibitem [{\citenamefont {Friedman}(2000)}]{friedman2000greedy}%
  \BibitemOpen
  \bibfield  {author} {\bibinfo {author} {\bibfnamefont {Jerome~H.}\ \bibnamefont {Friedman}},\ }\bibfield  {title} {\enquote {\bibinfo {title} {Greedy function approximation: A gradient boosting machine},}\ }\href@noop {} {\bibfield  {journal} {\bibinfo  {journal} {Annals of Statistics}\ }\textbf {\bibinfo {volume} {29}},\ \bibinfo {pages} {1189--1232} (\bibinfo {year} {2000})}\BibitemShut {NoStop}%
\bibitem [{\citenamefont {Finke}\ \emph {et~al.}(2024)\citenamefont {Finke}, \citenamefont {Hein}, \citenamefont {Kasieczka}, \citenamefont {Kr\"amer}, \citenamefont {M\"uck}, \citenamefont {Prangchaikul}, \citenamefont {Quadfasel}, \citenamefont {Shih},\ and\ \citenamefont {Sommerhalder}}]{Finke:2023ltw}%
  \BibitemOpen
  \bibfield  {author} {\bibinfo {author} {\bibfnamefont {Thorben}\ \bibnamefont {Finke}}, \bibinfo {author} {\bibfnamefont {Marie}\ \bibnamefont {Hein}}, \bibinfo {author} {\bibfnamefont {Gregor}\ \bibnamefont {Kasieczka}}, \bibinfo {author} {\bibfnamefont {Michael}\ \bibnamefont {Kr\"amer}}, \bibinfo {author} {\bibfnamefont {Alexander}\ \bibnamefont {M\"uck}}, \bibinfo {author} {\bibfnamefont {Parada}\ \bibnamefont {Prangchaikul}}, \bibinfo {author} {\bibfnamefont {Tobias}\ \bibnamefont {Quadfasel}}, \bibinfo {author} {\bibfnamefont {David}\ \bibnamefont {Shih}}, \ and\ \bibinfo {author} {\bibfnamefont {Manuel}\ \bibnamefont {Sommerhalder}},\ }\bibfield  {title} {\enquote {\bibinfo {title} {{Tree-based algorithms for weakly supervised anomaly detection}},}\ }\href {\doibase 10.1103/PhysRevD.109.034033} {\bibfield  {journal} {\bibinfo  {journal} {Phys. Rev. D}\ }\textbf {\bibinfo {volume} {109}},\ \bibinfo {pages} {034033} (\bibinfo {year} {2024})},\ \Eprint {http://arxiv.org/abs/2309.13111} {arXiv:2309.13111
  [hep-ph]} \BibitemShut {NoStop}%
\bibitem [{\citenamefont {Freytsis}\ \emph {et~al.}(2024)\citenamefont {Freytsis}, \citenamefont {Perelstein},\ and\ \citenamefont {San}}]{Freytsis:2023cjr}%
  \BibitemOpen
  \bibfield  {author} {\bibinfo {author} {\bibfnamefont {Marat}\ \bibnamefont {Freytsis}}, \bibinfo {author} {\bibfnamefont {Maxim}\ \bibnamefont {Perelstein}}, \ and\ \bibinfo {author} {\bibfnamefont {Yik~Chuen}\ \bibnamefont {San}},\ }\bibfield  {title} {\enquote {\bibinfo {title} {{Anomaly detection in the presence of irrelevant features}},}\ }\href {\doibase 10.1007/JHEP02(2024)220} {\bibfield  {journal} {\bibinfo  {journal} {JHEP}\ }\textbf {\bibinfo {volume} {02}},\ \bibinfo {pages} {220} (\bibinfo {year} {2024})},\ \Eprint {http://arxiv.org/abs/2310.13057} {arXiv:2310.13057 [hep-ph]} \BibitemShut {NoStop}%
\bibitem [{\citenamefont {Neyman}\ and\ \citenamefont {Pearson}(1933)}]{NP}%
  \BibitemOpen
  \bibfield  {author} {\bibinfo {author} {\bibfnamefont {J.}~\bibnamefont {Neyman}}\ and\ \bibinfo {author} {\bibfnamefont {E.~S.}\ \bibnamefont {Pearson}},\ }\bibfield  {title} {\enquote {\bibinfo {title} {On the problem of the most efficient tests of statistical hypotheses},}\ }\href {http://www.jstor.org/stable/91247} {\bibfield  {journal} {\bibinfo  {journal} {Philosophical Transactions of the Royal Society of London. Series A, Containing Papers of a Mathematical or Physical Character}\ }\textbf {\bibinfo {volume} {231}},\ \bibinfo {pages} {289--337} (\bibinfo {year} {1933})}\BibitemShut {NoStop}%
\bibitem [{\citenamefont {Metodiev}\ \emph {et~al.}(2017)\citenamefont {Metodiev}, \citenamefont {Nachman},\ and\ \citenamefont {Thaler}}]{Metodiev:2017vrx}%
  \BibitemOpen
  \bibfield  {author} {\bibinfo {author} {\bibfnamefont {Eric~M.}\ \bibnamefont {Metodiev}}, \bibinfo {author} {\bibfnamefont {Benjamin}\ \bibnamefont {Nachman}}, \ and\ \bibinfo {author} {\bibfnamefont {Jesse}\ \bibnamefont {Thaler}},\ }\bibfield  {title} {\enquote {\bibinfo {title} {{Classification without labels: Learning from mixed samples in high energy physics}},}\ }\href {\doibase 10.1007/JHEP10(2017)174} {\bibfield  {journal} {\bibinfo  {journal} {JHEP}\ }\textbf {\bibinfo {volume} {10}},\ \bibinfo {pages} {174} (\bibinfo {year} {2017})},\ \Eprint {http://arxiv.org/abs/1708.02949} {arXiv:1708.02949 [hep-ph]} \BibitemShut {NoStop}%
\bibitem [{\citenamefont {Ojala}\ and\ \citenamefont {Garriga}(2009)}]{5360332}%
  \BibitemOpen
  \bibfield  {author} {\bibinfo {author} {\bibfnamefont {Markus}\ \bibnamefont {Ojala}}\ and\ \bibinfo {author} {\bibfnamefont {Gemma~C.}\ \bibnamefont {Garriga}},\ }\bibfield  {title} {\enquote {\bibinfo {title} {Permutation tests for studying classifier performance},}\ }in\ \href {\doibase 10.1109/ICDM.2009.108} {\emph {\bibinfo {booktitle} {2009 Ninth IEEE International Conference on Data Mining}}}\ (\bibinfo {year} {2009})\ pp.\ \bibinfo {pages} {908--913}\BibitemShut {NoStop}%
\bibitem [{\citenamefont {Cowan}\ \emph {et~al.}(2011)\citenamefont {Cowan}, \citenamefont {Cranmer}, \citenamefont {Gross},\ and\ \citenamefont {Vitells}}]{Cowan:2010js}%
  \BibitemOpen
  \bibfield  {author} {\bibinfo {author} {\bibfnamefont {Glen}\ \bibnamefont {Cowan}}, \bibinfo {author} {\bibfnamefont {Kyle}\ \bibnamefont {Cranmer}}, \bibinfo {author} {\bibfnamefont {Eilam}\ \bibnamefont {Gross}}, \ and\ \bibinfo {author} {\bibfnamefont {Ofer}\ \bibnamefont {Vitells}},\ }\bibfield  {title} {\enquote {\bibinfo {title} {{Asymptotic formulae for likelihood-based tests of new physics}},}\ }\href {\doibase 10.1140/epjc/s10052-011-1554-0} {\bibfield  {journal} {\bibinfo  {journal} {Eur. Phys. J. C}\ }\textbf {\bibinfo {volume} {71}},\ \bibinfo {pages} {1554} (\bibinfo {year} {2011})},\ \bibinfo {note} {[Erratum: Eur.Phys.J.C 73, 2501 (2013)]},\ \Eprint {http://arxiv.org/abs/1007.1727} {arXiv:1007.1727 [physics.data-an]} \BibitemShut {NoStop}%
\bibitem [{\citenamefont {Freytsis}\ \emph {et~al.}(2010)\citenamefont {Freytsis}, \citenamefont {Ovanesyan},\ and\ \citenamefont {Thaler}}]{Freytsis:2009bh}%
  \BibitemOpen
  \bibfield  {author} {\bibinfo {author} {\bibfnamefont {Marat}\ \bibnamefont {Freytsis}}, \bibinfo {author} {\bibfnamefont {Grigory}\ \bibnamefont {Ovanesyan}}, \ and\ \bibinfo {author} {\bibfnamefont {Jesse}\ \bibnamefont {Thaler}},\ }\bibfield  {title} {\enquote {\bibinfo {title} {{Dark Force Detection in Low Energy e-p Collisions}},}\ }\href {\doibase 10.1007/JHEP01(2010)111} {\bibfield  {journal} {\bibinfo  {journal} {JHEP}\ }\textbf {\bibinfo {volume} {01}},\ \bibinfo {pages} {111} (\bibinfo {year} {2010})},\ \Eprint {http://arxiv.org/abs/0909.2862} {arXiv:0909.2862 [hep-ph]} \BibitemShut {NoStop}%
\bibitem [{\citenamefont {Kondo}(1988)}]{Kondo:1988yd}%
  \BibitemOpen
  \bibfield  {author} {\bibinfo {author} {\bibfnamefont {K.}~\bibnamefont {Kondo}},\ }\bibfield  {title} {\enquote {\bibinfo {title} {{Dynamical Likelihood Method for Reconstruction of Events With Missing Momentum. 1: Method and Toy Models}},}\ }\href {\doibase 10.1143/JPSJ.57.4126} {\bibfield  {journal} {\bibinfo  {journal} {J. Phys. Soc. Jap.}\ }\textbf {\bibinfo {volume} {57}},\ \bibinfo {pages} {4126--4140} (\bibinfo {year} {1988})}\BibitemShut {NoStop}%
\bibitem [{\citenamefont {Kondo}(1991)}]{Kondo:1991dw}%
  \BibitemOpen
  \bibfield  {author} {\bibinfo {author} {\bibfnamefont {K.}~\bibnamefont {Kondo}},\ }\bibfield  {title} {\enquote {\bibinfo {title} {{Dynamical likelihood method for reconstruction of events with missing momentum. 2: Mass spectra for 2 ---\ensuremath{>} 2 processes}},}\ }\href {\doibase 10.1143/JPSJ.60.836} {\bibfield  {journal} {\bibinfo  {journal} {J. Phys. Soc. Jap.}\ }\textbf {\bibinfo {volume} {60}},\ \bibinfo {pages} {836--844} (\bibinfo {year} {1991})}\BibitemShut {NoStop}%
\bibitem [{\citenamefont {Bohm}\ and\ \citenamefont {Zech}(2014)}]{Bohm:2013gla}%
  \BibitemOpen
  \bibfield  {author} {\bibinfo {author} {\bibfnamefont {G.}~\bibnamefont {Bohm}}\ and\ \bibinfo {author} {\bibfnamefont {G.}~\bibnamefont {Zech}},\ }\bibfield  {title} {\enquote {\bibinfo {title} {{Statistics of weighted Poisson events and its applications}},}\ }\href {\doibase 10.1016/j.nima.2014.02.021} {\bibfield  {journal} {\bibinfo  {journal} {Nucl. Instrum. Meth. A}\ }\textbf {\bibinfo {volume} {748}},\ \bibinfo {pages} {1--6} (\bibinfo {year} {2014})},\ \Eprint {http://arxiv.org/abs/1309.1287} {arXiv:1309.1287 [physics.data-an]} \BibitemShut {NoStop}%
\bibitem [{\citenamefont {Swartz}\ \emph {et~al.}(1990)\citenamefont {Swartz} \emph {et~al.}}]{Swartz:1990ki}%
  \BibitemOpen
  \bibfield  {author} {\bibinfo {author} {\bibfnamefont {Morris~L.}\ \bibnamefont {Swartz}} \emph {et~al.},\ }\bibfield  {title} {\enquote {\bibinfo {title} {{A Search for Doubly Charged Higgs Scalars in $Z$ Decay}},}\ }\href {\doibase 10.1103/PhysRevLett.64.2877} {\bibfield  {journal} {\bibinfo  {journal} {Phys. Rev. Lett.}\ }\textbf {\bibinfo {volume} {64}},\ \bibinfo {pages} {2877--2880} (\bibinfo {year} {1990})}\BibitemShut {NoStop}%
\bibitem [{\citenamefont {Atag}\ and\ \citenamefont {Ozansoy}(2003)}]{Atag:2003wk}%
  \BibitemOpen
  \bibfield  {author} {\bibinfo {author} {\bibfnamefont {S.}~\bibnamefont {Atag}}\ and\ \bibinfo {author} {\bibfnamefont {K.~O.}\ \bibnamefont {Ozansoy}},\ }\bibfield  {title} {\enquote {\bibinfo {title} {{Realistic constraints on the doubly charged bilepton couplings from Bhabha scattering with LEP data}},}\ }\href {\doibase 10.1103/PhysRevD.68.093008} {\bibfield  {journal} {\bibinfo  {journal} {Phys. Rev. D}\ }\textbf {\bibinfo {volume} {68}},\ \bibinfo {pages} {093008} (\bibinfo {year} {2003})},\ \Eprint {http://arxiv.org/abs/hep-ph/0310046} {arXiv:hep-ph/0310046} \BibitemShut {NoStop}%
\bibitem [{\citenamefont {Chen}\ and\ \citenamefont {Guestrin}(2016)}]{Chen_2016}%
  \BibitemOpen
  \bibfield  {author} {\bibinfo {author} {\bibfnamefont {Tianqi}\ \bibnamefont {Chen}}\ and\ \bibinfo {author} {\bibfnamefont {Carlos}\ \bibnamefont {Guestrin}},\ }\bibfield  {title} {\enquote {\bibinfo {title} {Xgboost: A scalable tree boosting system},}\ }in\ \href {\doibase 10.1145/2939672.2939785} {\emph {\bibinfo {booktitle} {Proceedings of the 22nd ACM SIGKDD International Conference on Knowledge Discovery and Data Mining}}},\ \bibinfo {series and number} {KDD ’16}\ (\bibinfo  {publisher} {ACM},\ \bibinfo {year} {2016})\ p.\ \bibinfo {pages} {785–794}\BibitemShut {NoStop}%
\bibitem [{\citenamefont {Ansel}\ \emph {et~al.}(2024)\citenamefont {Ansel}, \citenamefont {Yang}, \citenamefont {He}, \citenamefont {Gimelshein}, \citenamefont {Jain}, \citenamefont {Voznesensky}, \citenamefont {Bao}, \citenamefont {Bell}, \citenamefont {Berard}, \citenamefont {Burovski}, \citenamefont {Chauhan}, \citenamefont {Chourdia}, \citenamefont {Constable}, \citenamefont {Desmaison}, \citenamefont {DeVito}, \citenamefont {Ellison}, \citenamefont {Feng}, \citenamefont {Gong}, \citenamefont {Gschwind}, \citenamefont {Hirsh}, \citenamefont {Huang}, \citenamefont {Kalambarkar}, \citenamefont {Kirsch}, \citenamefont {Lazos}, \citenamefont {Lezcano}, \citenamefont {Liang}, \citenamefont {Liang}, \citenamefont {Lu}, \citenamefont {Luk}, \citenamefont {Maher}, \citenamefont {Pan}, \citenamefont {Puhrsch}, \citenamefont {Reso}, \citenamefont {Saroufim}, \citenamefont {Siraichi}, \citenamefont {Suk}, \citenamefont {Suo}, \citenamefont {Tillet}, \citenamefont {Wang}, \citenamefont {Wang}, \citenamefont {Wen},
  \citenamefont {Zhang}, \citenamefont {Zhao}, \citenamefont {Zhou}, \citenamefont {Zou}, \citenamefont {Mathews}, \citenamefont {Chanan}, \citenamefont {Wu},\ and\ \citenamefont {Chintala}}]{Ansel_PyTorch_2_Faster_2024}%
  \BibitemOpen
  \bibfield  {author} {\bibinfo {author} {\bibfnamefont {Jason}\ \bibnamefont {Ansel}}, \bibinfo {author} {\bibfnamefont {Edward}\ \bibnamefont {Yang}}, \bibinfo {author} {\bibfnamefont {Horace}\ \bibnamefont {He}}, \bibinfo {author} {\bibfnamefont {Natalia}\ \bibnamefont {Gimelshein}}, \bibinfo {author} {\bibfnamefont {Animesh}\ \bibnamefont {Jain}}, \bibinfo {author} {\bibfnamefont {Michael}\ \bibnamefont {Voznesensky}}, \bibinfo {author} {\bibfnamefont {Bin}\ \bibnamefont {Bao}}, \bibinfo {author} {\bibfnamefont {Peter}\ \bibnamefont {Bell}}, \bibinfo {author} {\bibfnamefont {David}\ \bibnamefont {Berard}}, \bibinfo {author} {\bibfnamefont {Evgeni}\ \bibnamefont {Burovski}}, \bibinfo {author} {\bibfnamefont {Geeta}\ \bibnamefont {Chauhan}}, \bibinfo {author} {\bibfnamefont {Anjali}\ \bibnamefont {Chourdia}}, \bibinfo {author} {\bibfnamefont {Will}\ \bibnamefont {Constable}}, \bibinfo {author} {\bibfnamefont {Alban}\ \bibnamefont {Desmaison}}, \bibinfo {author} {\bibfnamefont {Zachary}\ \bibnamefont
  {DeVito}}, \bibinfo {author} {\bibfnamefont {Elias}\ \bibnamefont {Ellison}}, \bibinfo {author} {\bibfnamefont {Will}\ \bibnamefont {Feng}}, \bibinfo {author} {\bibfnamefont {Jiong}\ \bibnamefont {Gong}}, \bibinfo {author} {\bibfnamefont {Michael}\ \bibnamefont {Gschwind}}, \bibinfo {author} {\bibfnamefont {Brian}\ \bibnamefont {Hirsh}}, \bibinfo {author} {\bibfnamefont {Sherlock}\ \bibnamefont {Huang}}, \bibinfo {author} {\bibfnamefont {Kshiteej}\ \bibnamefont {Kalambarkar}}, \bibinfo {author} {\bibfnamefont {Laurent}\ \bibnamefont {Kirsch}}, \bibinfo {author} {\bibfnamefont {Michael}\ \bibnamefont {Lazos}}, \bibinfo {author} {\bibfnamefont {Mario}\ \bibnamefont {Lezcano}}, \bibinfo {author} {\bibfnamefont {Yanbo}\ \bibnamefont {Liang}}, \bibinfo {author} {\bibfnamefont {Jason}\ \bibnamefont {Liang}}, \bibinfo {author} {\bibfnamefont {Yinghai}\ \bibnamefont {Lu}}, \bibinfo {author} {\bibfnamefont {CK}~\bibnamefont {Luk}}, \bibinfo {author} {\bibfnamefont {Bert}\ \bibnamefont {Maher}}, \bibinfo {author}
  {\bibfnamefont {Yunjie}\ \bibnamefont {Pan}}, \bibinfo {author} {\bibfnamefont {Christian}\ \bibnamefont {Puhrsch}}, \bibinfo {author} {\bibfnamefont {Matthias}\ \bibnamefont {Reso}}, \bibinfo {author} {\bibfnamefont {Mark}\ \bibnamefont {Saroufim}}, \bibinfo {author} {\bibfnamefont {Marcos~Yukio}\ \bibnamefont {Siraichi}}, \bibinfo {author} {\bibfnamefont {Helen}\ \bibnamefont {Suk}}, \bibinfo {author} {\bibfnamefont {Michael}\ \bibnamefont {Suo}}, \bibinfo {author} {\bibfnamefont {Phil}\ \bibnamefont {Tillet}}, \bibinfo {author} {\bibfnamefont {Eikan}\ \bibnamefont {Wang}}, \bibinfo {author} {\bibfnamefont {Xiaodong}\ \bibnamefont {Wang}}, \bibinfo {author} {\bibfnamefont {William}\ \bibnamefont {Wen}}, \bibinfo {author} {\bibfnamefont {Shunting}\ \bibnamefont {Zhang}}, \bibinfo {author} {\bibfnamefont {Xu}~\bibnamefont {Zhao}}, \bibinfo {author} {\bibfnamefont {Keren}\ \bibnamefont {Zhou}}, \bibinfo {author} {\bibfnamefont {Richard}\ \bibnamefont {Zou}}, \bibinfo {author} {\bibfnamefont {Ajit}\
  \bibnamefont {Mathews}}, \bibinfo {author} {\bibfnamefont {Gregory}\ \bibnamefont {Chanan}}, \bibinfo {author} {\bibfnamefont {Peng}\ \bibnamefont {Wu}}, \ and\ \bibinfo {author} {\bibfnamefont {Soumith}\ \bibnamefont {Chintala}},\ }\bibfield  {title} {\enquote {\bibinfo {title} {{PyTorch 2: Faster Machine Learning Through Dynamic Python Bytecode Transformation and Graph Compilation}},}\ }in\ \href {\doibase 10.1145/3620665.3640366} {\emph {\bibinfo {booktitle} {29th ACM International Conference on Architectural Support for Programming Languages and Operating Systems, Volume 2 (ASPLOS '24)}}}\ (\bibinfo  {publisher} {ACM},\ \bibinfo {year} {2024})\BibitemShut {NoStop}%
\end{thebibliography}%

\end{document}